\documentclass[twocolumn,showpacs,prb]{revtex4}

\usepackage{graphicx}%
\usepackage{dcolumn}
\usepackage{amsmath}
\usepackage{amssymb}

\makeatletter
\def\btt#1{\texttt{\@backslashchar#1}}%
\DeclareRobustCommand\bblash{\btt{\@backslashchar}}%
\makeatother

\begin{document}

\title[Short Title]{Anomalous Josephson effect in $\boldsymbol{p}$-wave dirty junctions}

\author{Yasuhiro Asano$^1$, Yukio Tanaka$^2$ and Satoshi Kashiwaya$^3$}
\affiliation{
$^1$Department of Applied Physics, Hokkaido University, Sapporo 060-8628, Japan\\
$^2$Department of Applied Physics, Nagoya University, Nagoya 464-8603, Japan\\
$^2$CREST, Japan Science and Technology Corporation (JST)
Nagoya, 464-8603, Japan 
\\$^3$National Institute of Advanced Industrial Science and Technology, 
Tsukuba 305-8568, Japan
}%

\date{\today}

\begin{abstract}
The Josephson effect in $p$-wave superconductor / diffusive normal metal / $p$-wave superconductor
junctions is studied theoretically.
Amplitudes of Josephson currents are several orders of magnitude larger than those in $s$-wave 
junctions. Current-phase ($J$-$\varphi$) relations in low temperatures are close to those in 
ballistic junctions such as $J\propto\sin(\varphi/2)$ and $J\propto\varphi$ even in the presence
of random impurity potentials.
A cooperative effect between the midgap Andreev resonant 
states and the proximity effect causes such anomalous properties and is 
a character of the spin-triplet superconductor junctions.
\end{abstract}

\maketitle
The {\em internal} $\pi$-phase shift (sign change) of pair potentials is essential for 
unconventional superconductivity and 
is the source of the midgap Andreev resonant state (MARS)~\cite{buchholts,hara,hu}.
It is now known that the MARS  
is responsible for anomalous transport properties in superconducting
junctions~\cite{tanaka0}. In normal metal/superconductor junctions, 
transport properties are affected also by the
proximity effect which is interpreted in terms of diffusion of Cooper pairs into normal metals.
In what follows, we assume that normal metals are in the diffusive transport regime
due to impurity scatterings. 
Recent theoretical studies have revealed sensitivity 
of the proximity effect to the {\em internal} phase of pair potentials~\cite{ya01-2,yt03-1}.
In normal metals attached to unconventional superconductors,
Cooper pairs have a sign degree of freedom reflecting 
the $\pi$-phase shift of pair potentials.
Suppression of the proximity effect is usually expected 
because wave function of a Cooper pair originated from the positive part of pair potentials 
cancel that originated from the negative part~\cite{ya01-2,yt03-1}. 
Two of us, however, discussed anomalous enhancement of 
the zero-bias tunneling conductance due to the proximity effect
in a presence of the MARS~\cite{yt04,yt05}. 

In superconductor / nomal metal / superconductor junctions, another phase degree of 
freedom affects quantum transport. Namely, the {\em external} phase difference 
 across the junctions $\varphi$ drives Josephson currents. 
An importance of studying the Josephson effect is growing these days
because quantum interference devices consisting of Josephson junctions can be basis of 
future technologies.
In fact, a recent experiment has tried to apply high-$T_c$ 
superconductors to coherent devices~\cite{bauch}. 
In unconventional junctions, the MARS is considered to have the phase degree of freedom.
When MARS's are formed at the two junction interfaces, the external phase may modify 
interference effects between the two MARS's and Josephson currents. 
The research in this direction can shed new light on quantum transport 
in unconventional superconductors.

In this paper, we theoretically study Josephson currents between 
two $p$-wave superconductors through normal metals
by solving the Bogoliubov-de Gennes equation 
using the recursive Green function method~\cite{furusaki,ya01-1}.
We show that amplitudes of Josephson currents in the $p$-wave junctions are much larger than those 
in the $s$-wave junctions when transmission probabilities of junction interfaces are small.
The local density of states in normal metals has a zero-energy peak reflecting 
anomalous diffusion of the MARS's into a normal metal and that 
spatial profiles of the zero-energy peak depend strongly on $\varphi$. 
As a consequence, current-phase ($J$-$\varphi$) relations remarkably deviate from
the sinusoidal function in low temperatures and are close to those in ballistic junctions 
such as $J\propto \sin(\varphi/2)$ and $J\propto \varphi$~\cite{ko2,ishii,bardeen,golubov}.
The resonant tunneling through the MARS {\em in} normal metals is responsible for such 
unusual Josephson effect.
The obtained results imply high potentials of spin-triplet superconducting junctions 
as coherent devices.

\begin{figure}[thb]
\begin{center}
\includegraphics[width=8.0cm]{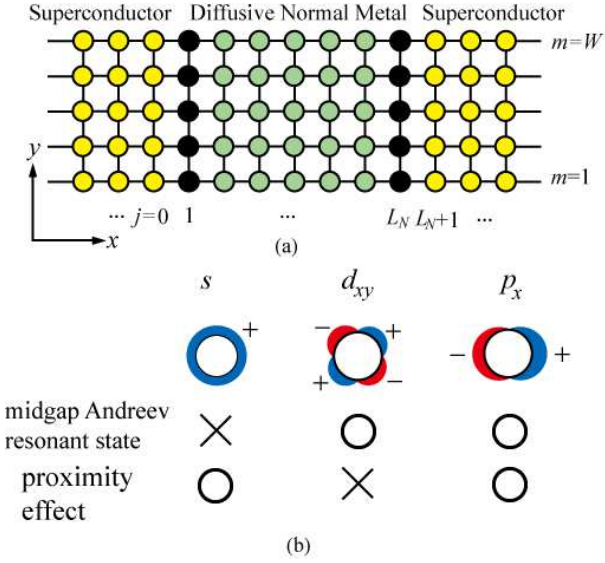}
\end{center}
\caption{ A schematic figure of a Josephson junction on the tight-binding
lattice is shown in (a). In (b), we illustrate the pair potentials in 
momentum space, where open circles represent the Fermi surface.
The pair potentials are classified into three groups by the presence 
or absence of the two interference effects.
}
\label{fig1}
\end{figure}

We consider three pairing symmetries on two dimensional superconductors:
(i) $\Delta_{\boldsymbol{k}}= \Delta_0$ for $s$-wave, (ii) $\Delta_02\bar{k}_x \bar{k}_y$ for $d_{xy}$-wave, and (iii) $\Delta_0 \bar{k}_x$ for $p_x$-wave symmetries. 
Here $\Delta_0$ is the maximum amplitude of pair potentials at the zero temperature, 
$\bar{k}_x=k_x/k_F$ and $\bar{k}_y=k_y/k_F$ are normalized wave numbers 
on the Fermi surface in the $x$ and $y$ directions, respectively. 
Josephson currents are parallel to the $x$ direction and junction interfaces 
are parallel to the $y$ direction as shown in Fig.~\ref{fig1}(a). 
The pair potentials in momentum space are illustrated in Fig.~\ref{fig1}(b).
An interference of a quasiparticle enables formation of the MARS 
at a junction interface when $\Delta_{k_x,k_y} \Delta_{-k_x,k_y} < 0$~\cite{tanaka0,ya04}.
The pair potentials in the $d_{xy}$- and $p_x$-symmetries satisfy the relation
for all wave numbers.
The absence of the proximity effect in normal metals is described 
by a relation~\cite{ya01-2,yt03-1} 
$\Delta_{k_x,k_y}=-\Delta_{k_x,-k_y}$.
The pair potential in the $d_{xy}$-symmetry satisfies the relation.
Thus the proximity effect is expected in both the $s$- and $p_x$-wave symmetries.
In Fig.~\ref{fig1}(b), we classify the pairing symmetries into three groups
by the presence ($\circ$) or absence ($\times$) of the two interference 
effects~\cite{yt04,yt05}.
Within $p$-wave symmetries, we pay special attention to the $p_x$-wave symmetry because 
the proximity effect and MARS are present at the same time.
On the other hand in the $p_y$-wave symmetry, neither is present~\cite{yt04}.

Let us consider Josephson 
junctions on the two-dimensional tight-binding model as shown in 
Fig.~\ref{fig1}(a). A vector $\boldsymbol{r}=j {\boldsymbol{x}}+ m{\boldsymbol{y}}$
points a lattice site, where ${\boldsymbol{x}}$ and ${\boldsymbol{y}}$ are the unit 
vectors in the $x$ and $y$ directions, respectively.
The junction consists of three regions: a normal metal (i.e., $1 \leq j \leq L_N$) 
and two superconductors (i.e., $-\infty \leq j \leq 0$ and $L_N+1 \leq j \leq \infty$).
In the $y$ direction, the number of lattice sites is $W$ and we assume the periodic boundary 
condition. 
Electronic states in superconducting junctions are described by the mean-field Hamiltonian
\begin{align}
H_{\textrm{BCS}}=& \frac{1}{2}\sum_{\boldsymbol{r},\boldsymbol{r}'} 
 \left[ \tilde{c}_{\boldsymbol{r}}^\dagger\;  \hat{h}_{\boldsymbol{r},\boldsymbol{r}'} 
  \; \tilde{c}_{\boldsymbol{r}'}^{ }  -  
\overline{\tilde{c}_{\boldsymbol{r}}}\;  \hat{h}^\ast_{\boldsymbol{r},\boldsymbol{r}'}  
 \; \overline{ \tilde{c}_{\boldsymbol{r}'}^\dagger }\; \right] \nonumber \\
 + \frac{1}{2} \sum_{\boldsymbol{r},\boldsymbol{r}' \in \textrm{S}}&
 \left[ \tilde{c}_{\boldsymbol{r}}^\dagger \;
\hat{\Delta}_{\boldsymbol{r},\boldsymbol{r}'}\;
\overline{\tilde{c}_{\boldsymbol{r}'}^\dagger}
- \overline{\tilde{c}_{\boldsymbol{r}}}\; 
\hat{\Delta}^\ast_{\boldsymbol{r},\boldsymbol{r}'}\;
\tilde{c}_{\boldsymbol{r}'} \right], \label{bcs}
\end{align}
\begin{align}
\hat{h}_{\boldsymbol{r},\boldsymbol{r}'}=& \left[-t \delta_{|\boldsymbol{r}-\boldsymbol{r}'|,1}
+ (\epsilon_{\boldsymbol{r}}- \mu+4t)\delta_{\boldsymbol{r},\boldsymbol{r}'}\right]\hat{\sigma}_0 \nonumber \\
+ &\boldsymbol{V}(\boldsymbol{r}\cdot\hat{\boldsymbol{\sigma}}\\
\hat{\Delta}_{\boldsymbol{r},\boldsymbol{r}'}=& 
i \Delta \hat{\sigma}_2, \\
\tilde{c}_{\boldsymbol{r}}=&\left( \begin{array}{c} c_{\boldsymbol{r},\uparrow} \\
c_{\boldsymbol{r},\downarrow}\end{array}\right),
\end{align}
where $c_{\boldsymbol{r},\sigma}^{\dagger}$ ($c_{\boldsymbol{r},\sigma}^{ }$) 
is the creation (annihilation) operator of an electron at $\boldsymbol{r}$ with 
spin $\sigma =$ ( $\uparrow$ or $\downarrow$ ) and $\overline{ \tilde{c}}$ means
the transpose of $\tilde{c}$.
The hopping integral $t$ is considered among nearest neighbor sites.
We assume that $t$ and the Fermi energy $\mu$
are common in superconductors and a normal metal.
In a normal metal, on-site potentials are given randomly in the range of 
$-V_I/2 \leq \epsilon_{\boldsymbol{r}} \leq V_I/2$.
We introduce insulating barriers at $j=1$ and $L_N$, where 
$\epsilon_{\boldsymbol{r}}$ is given by $V_B$.
Two superconductors in which $\epsilon_{\boldsymbol{r}}$ are taken to be zero are 
identical to each other. 
In the $p_x$-wave symmetry, a spin vector of Cooper pairs $\boldsymbol{d}$ points 
the $z$ direction. The arguments below do not depend on directions of $\boldsymbol{d}$. 
The Hamiltonian is diagonalized by the Bogoliubov transformation and the
Bogoliubov-de Gennes equation is numerically solved by the recursive Green function
 method~\cite{furusaki,ya01-1}. 
Josephson currents are given by
\begin{align}
J = -i et T
 \sum_{\omega_n} {\rm Tr}
\left[  \hat{G}_{\omega_n}(\boldsymbol{r}', \boldsymbol{r}) 
 - \hat{G}_{\omega_n}(\boldsymbol{r}, \boldsymbol{r}')
\right] \label{jq}
\end{align}
with $\boldsymbol{r}'=\boldsymbol{r}+\boldsymbol{x}$,
where $\hat{G}_{\omega_n}$ is the Green function and 
$\omega_n = (2n+1)\pi T$ is the Matsubara frequency 
with $n$ and $T$ being an integer and a temperature, respectively. 
In Eq.~(\ref{jq}), Tr means the trace in the 
Nambu space and the summation over $m$.
In this paper, the unit of $\hbar=k_B=1$ is used with $k_B$ being the Boltzmann constant.  
Local density of states is also calculated from 
$N(E,j) = - \textrm{Im Tr} \; \hat{G}_{E+i\gamma}(\boldsymbol{r},\boldsymbol{r})/\pi$,
where $E$ is measured from the Fermi energy and $\gamma$ is a small imaginary part.
Throughout this paper, we fix parameters as $L_N=70$, $W=25$, $\mu=2t$, and $V_I=2t$.
Under these parameters, normal metals are in the diffusive transport regime, 
where the mean free path in normal metals is estimated about $\ell\sim 6$ lattice constants 
and the Thouless energy $E_{\textrm{th}}$ is calculated to be $1.6\times 10^{-3}t$. 
Results discussed below are qualitatively insensitive to these parameters.

\begin{figure}[thb]
\begin{center}
\includegraphics[width=9.5cm]{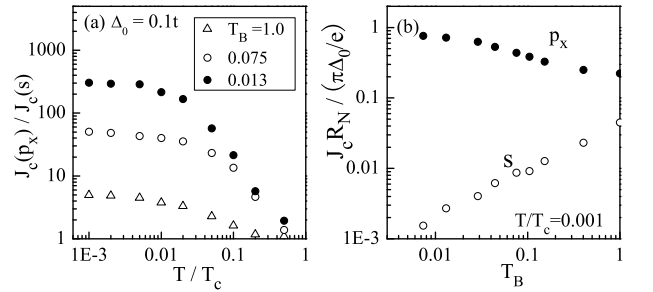}
\end{center}
\caption{ The maximum amplitudes of Josephson currents in the $p_x$-wave symmetry $J_c(p_x)$
are compared with those in the $s$-wave symmetry $J_c(s)$ in (a), where $T_B$
is the transmission probability of potential barriers in the normal states.
In (b), $J_c(p_x)$ and $J_c(s)$ are plotted as a function of $T_B$ at $T=0.001T_c$.
 }
\label{amp}
\end{figure}

At first we show that the maximum amplitudes of Josephson currents in the $p_x$-wave
symmetry $J_c(p_x)$ become much larger than those in the $s$-wave $J_c(s)$.
In Fig.~\ref{amp}(a), ratios $J_c(p_x)/J_c(s)$ are plotted as a function of 
temperatures for $\Delta_0=0.1t$.
Here we choose several values of the barrier potentials $V_B$ at $j=1$ and $L_N$. 
The resulting normal transmission probabilities of the barrier
$T_B$ are 1.0, 0.075 and 0.013 for $V_B/t =0$, 6 and 15, respectively.
 The ratios $J_c(p_x)/J_c(s)$ increase with decreasing $T$ and amazingly 
become more than 100 in low temperatures for small $T_B$.
The amplitudes of Josephson currents in the $p_x$-wave junctions are much 
larger than those in the $s$-wave junctions.
In Fig.~\ref{amp}(b), $J_cR_N$ normalized by $\pi\Delta_0/e$ is plotted
as a function of $T_B$ at $T=0.001T_c$, where $R_N$ is the normal resistance of junctions. 
The results show that $J_c R_N$ in the $s$-wave decreases with decreasing
$T_B$, whereas that in the $p_x$-wave increases. 

\begin{figure}[thb]
\begin{center}
\includegraphics[width=9.5cm]{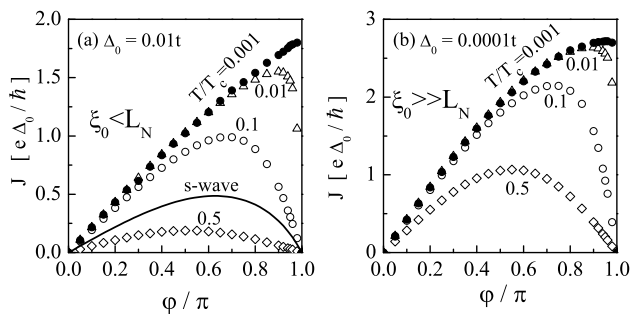}
\end{center}
\caption{ Current-phase relations for the $p_x$-wave 
symmetries are shown for several temperatures at $V_B=0$,
where $\Delta_0=0.01t$ in (a) and  $\Delta_0=0.0001t$ in (b).
For comparison, results in the $s$-wave junctions at $T=0.001T_c$ 
is shown with a solid line in (a),
where the amplitude of Josephson current is multiplied by 5.
 }
\label{cpr}
\end{figure}

We next focus on current-phase relations of the Josephson effect.
In Fig.~\ref{cpr}, Josephson currents are plotted as a function of $\varphi$
 for the $p_x$-wave symmetries at $V_B=0$. 
Parameters are chosen as $\Delta_0=0.01t$ and $0.0001t$ in (a) and (b), respectively.
The current-phase relations are almost sinusoidal function in a high temperature at $T=0.5T_c$.
At $T=0.001T_c$, however, 
the current-phase relations are close to $J\propto \varphi$ and 
$J\propto\sin(\varphi/2)$ in (a) and (b), respectively.
These are characteristic current-phase relations in ballistic Josephson junctions in the $s$-wave
symmetry~\cite{ishii,bardeen,ko2}.
We have confirmed that these current-phase relations remain even in the presence of potential 
barriers (i.e., $V_B\neq 0$).

The results imply large contributions of the multiple Andreev reflection in low temperatures. 
In general, Josephson currents can be decomposed into a series of 
$J=\sum_{n=1}^\infty J_n \sin(n\varphi)$,
where $J_n$ for $n \geq 2$ represent contributions of the multiple Andreev reflection.
Roughly speaking, $J_n$ is proportional to $\left\{T_N\right\}^n$ with $T_N$ being 
the transmission probability of a quasiparticle from the left superconductor to
the right superconductor through the normal segment (including two barriers and a normal metal).
Thus the multiple Andreev reflection is 
negligible (i.e., $J_1 \gg J_2 \gg J_3 \ldots$) for $T_N \ll 1$.
On the other hand in the case of $T_N=1$, the multiple Andreev reflection 
leads to the deviation of current-phase relations from the sinusoidal function. 
It is noted at $T_N=1$ that
we obtain $J\propto \varphi$ and $J\propto\sin(\varphi/2)$ at the zero temperature 
for $L_N\gg \xi_0$ and $L_N\ll \xi_0$, respectively~\cite{ishii,bardeen,ko2}. 

In Fig.~\ref{cpr}(a), we also show the current-phase relations in the $s$-wave symmetry
at $T=0.001T_c$ with a solid line. 
The current-phase relation in the $s$-wave is described almost
by the sinusoidal function~\cite{zaikin} because impurity potentials in 
normal metals suppress $T_N$ and therefore the multiple Andreev reflection.
In the $p_x$-wave junctions, the coherence length $\xi_0$ are estimated about 50 lattice 
constants in (a) and 5000 in (b).
Thus $L_N > \xi_0$ and $L_N \ll \xi_0$ are satisfied in (a) and (b), respectively.
The current-phase relations such as 
$J\propto\sin(\varphi/2)$ in (b) and $J\propto \varphi$ in (a)
are universal properties of the $p_x$-wave junctions in low temperatures because they 
are independent of the strength of barrier potentials and the 
degree of disorder in normal metals. 
The calculated results in Fig.~\ref{cpr} indicate $T_N=1$ even in the presence of
impurity potentials. The large amplitudes of the Josephson
current in Fig.~\ref{amp} are also explained by $T_N=1$.

\begin{figure}[tbh]
\begin{center}
\includegraphics[width=8.5cm]{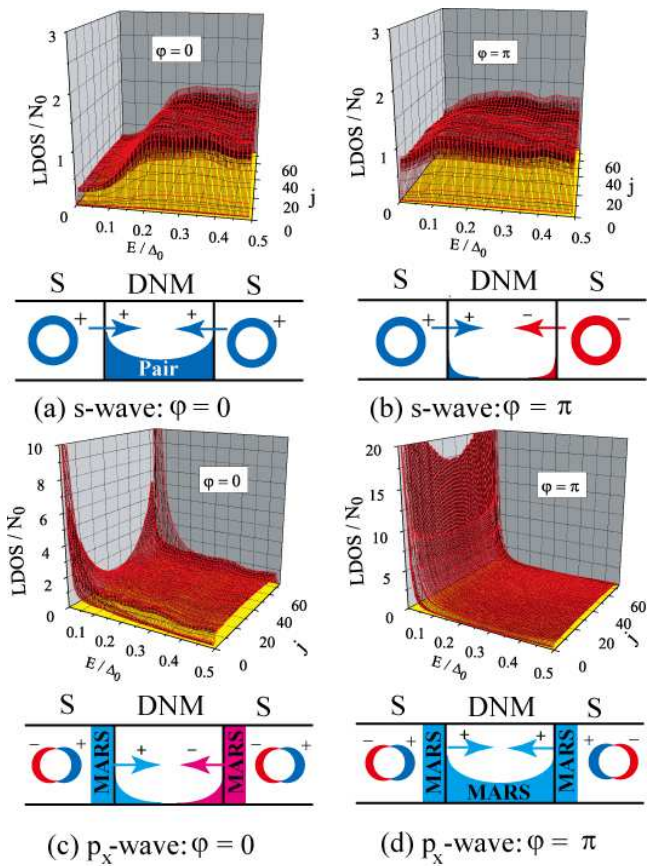}
\end{center}
\caption{ Local density of states (LDOS) in normal metals
($1 \leq j\leq L_N=70$) are shown for the
$s$-wave and $p_x$-wave symmetries. The left and right 
superconductors are attached at $j=0$ and $j= 71$, respectively.
Noting that $E_{th}$ is about $0.3 \Delta_0$.
In schematic pictures, DNM and S denote a diffusive normal metal and a superconductor,
respectively. 
The local density of states shown here are calculated
in the absence of Josephson currents. We have confirmed that the results at $\varphi=0.99\pi$
qualitatively shows the same behavior as those at $\varphi=\pi$. 
}
\label{dos}
\end{figure}

The calculated results in Figs.~\ref{amp} and \ref{cpr} show 
the specific properties of Josephson currents in the $p_x$-wave junctions. 
In what follows, we analyze quasiparticle states in normal metals 
to understand the origin of the anomalous Josephson effect.
In Fig.~\ref{dos}, we show the local density of states in normal metals 
for the $s$- and $p_x$-wave symmetries, 
where $\Delta_0=0.005t$, $\gamma=0.05\Delta_0$, 
and $N_0$ denotes the normal density of states.
At $\varphi=0$ in the $s$-wave junctions in (a), the local density of states for 
$E < E_{th}\sim 0.3 \Delta_0$ is suppressed because of the proximity effect. 
The suppression of the local density of states indicates the conversion of 
quasiparticles to Cooper pairs in normal metals.
At $\varphi=\pi$ in (b), the local density of states recovers its amplitude for $E < E_{th}$.
The wave function of Cooper pairs from the left superconductor and 
that from the right one cancel each other around $\varphi\sim\pi$ as schematically 
illustrated in a picture below the calculated results. 

The local density of states is drastically changed in the $p_x$-wave symmetry
as shown in (c) and (d). 
Zero-energy peaks whose width 
is determined by $\gamma$ can be seen, which
means formation of the midgap Andreev resonant state (MARS) in normal metals.
Although the MARS originally localizes at junction interfaces~\cite{tanaka0},
the MARS penetrates into normal metals in the presence of the proximity effect.
Spatial profiles of the local density of states depend remarkably on 
the external phase difference as shown in (c) and (d).
 At $\varphi=0$, the zero-energy peak disappears at 
the center of normal metals ($j\sim 35$) because 
wave function of the MARS from the left superconductors cancel out that from the right one
as shown schematically in a lower pannel in (c). 
 On the other hand in (d), wave functions of the MARS in the two superconductors 
have the same sign with each other. Thus the two MARS's can penetrate deeply 
into normal metals and the zero-energy peak can be seen everywhere.
We note that the penetration of the MARS is possible only when the proximity effect is present 
in normal metals.
In fact, we have confirmed that no zero-energy peak is found in normal metals
 in the $d_{xy}$-wave 
symmetry (results are not shown) and that the ensemble average of Josephson currents 
vanishes because the proximity effect is absent in normal metals~\cite{ya01-2}.
Fig.~\ref{dos} indicates that the proximity effect bridges the two MARS's in the two 
superconductors. Thus $T_N=1$ holds because of the resonant transmission 
through the MARS in normal metals. 
The Josephson effect specific to the $p_x$-wave symmetry discussed in Figs.~\ref{amp} and \ref{cpr} are a consequence of
the diffusion of the MARS {\em into} normal metals.

In summary, we found anomalous behaviors of Josephson currents in superconductor / 
 normal metals / superconductor junctions in the $p_x$-wave symmetry. 
The maximum amplitudes of Josephson currents $J_c$ in the $p_x$-wave 
junctions become much larger than those in the $s$-wave junctions.
It is known that large values of $J_c$ are desired in device applications because 
$J_cR_N$ limits operation speeds of Josephson devices. 
Current-phase relations in low temperatures are close to those in ballistic junctions 
such as $J\propto\sin(\varphi/2)$ and $J\propto \varphi$
independent of the strength of potential barriers at interfaces and the degree of 
disorder in normal metals.
The two the midgap Andreev resonant states penetrate deeply into normal metals, which causes 
the unusual Josephson effect in $p_x$-wave superconducting junctions.
The anomalous Josephson effect is a novel feature of phase-sensitive transport 
in spin-triplet superconducting junctions.

 This work has been partially supported by Grant-in-Aid for the 21st Century
COE program on "Topological Science and Technology" and "Frontiers of Computational Science" 
from the Ministry of Education, Culture, Sport, Science and Technology of Japan.

\end{document}